\documentstyle[multicol,amssymb,epsf,prl,aps]{revtex}
\begin{document}
\draft
\title{\large {\bf Nuclear Shell Effects near r-Process Path at N=82
in the Relativistic Hartree-Bogoliubov Theory}}
\author{Madan M. Sharma and Ameena R. Farhan}
\address{Physics Department, Kuwait University, Kuwait 13060}
\date{\today}
\maketitle

\begin{abstract}
Evolution of the shell structure of nuclei near the neutron drip line
is investigated in the Relativistic Hartree-Bogoliubov (RHB) theory.
By introducing the vector self-coupling of $\omega$ meson in the RHB theory,
we reproduce successfully the experimental data on the shell effects
about the waiting-point nucleus $^{80}$Zn. With this basis, it is shown
that the shell effects at $N=82$ in the inaccessible region of the
r-process path remain strong. In contrast, a quenching exhibited by
the HFB+SkP approach is shown to be incompatible with the available
data. Consequently, the neutrino-driven mechanism of the nucleosynthesis
is supported.
\end{abstract}

\begin{multicols}{2}

A knowledge of shell effects near the drip lines is important to 
discerning astrophysical scenario of nucleosynthesis \cite{Kratz.93}. 
The question whether the shell effects near the drip lines are strong 
or do quench has become crucial to understanding heavy nucleosynthesis. 
The $N=82$ nuclei at r-process path are assumed to play a significant
role in providing nuclear abundances about $A \sim 130$.
Since nuclei contributing to this peak are extremely neutron-rich
and are not accessible experimentally, it has not been possible to
ascertain the nature of the shell effects in the vicinity of the 
neutron drip line. Due to the lack of experimental data, there prevail 
conflicting view points \cite{SLHR.94,DHNA.94} on the strength of the 
shell effects near the neutron drip line. There exist different 
scenarios regarding the r-process nuclear abundances such as the
one assisted by shell quenching \cite{Kratz.93}, or that of 
neutrino-driven post-processing \cite{Hax.97,Qian.97}.

In this letter, we examine how the shell effects evolve with isospin
in the region of the astrophysically important magic number
$N=82$ near the neutron drip line. Using the experimental data available
on the shell effects around the waiting-point nucleus 
$^{80}$Zn ($N=50$) as a benchmark, we will explore the shell 
effects near the r-process path in the framework of the Relativistic
Hartree-Bogoliubov (RHB) theory with the self-consistent
finite-range pairing. We have introduced the vector self-coupling of 
$\omega$ meson in addition to the non-linear scalar coupling of the 
$\sigma$-meson in the RHB theory. The high-precision experimental data 
about the stability line and a successful description of the ensuing shell 
effects \cite{SFM.00} constitutes our basis to predict the shell effects 
in the inaccessible region.   

The shell effects are known to manifest strongly in terms of the magic 
numbers. This is demonstrated by a prominent kink about the major magic 
numbers \cite{Bor.93} in the 2-neutron separation energies ($S_{2n}$) 
all over the periodic table. This implies that there exist large 
shell gaps at the magic numbers about the stability line.
The spin-orbit interaction is pivotal \cite{MGM.55} in creation of 
the magic numbers. In the Relativistic Mean Field (RMF) theory
\cite{Serot.86} the spin-orbit term arises as a result 
of the Dirac-Lorentz structure of nucleons. This has shown much 
usefulness in explaining properties which involve
shell effects such as anomalous isotope shifts in stable nuclei
\cite{SLR.93}. The form of the spin-orbit interaction in the RMF
theory has also been found to be advantageous over that in the 
non-relativistic approach \cite{SLKR.94}. 

For nuclei near a drip line, a coupling to the continuum is required.
A self-consistent  treatment of pairing is also desirable. The framework
of RHB theory provides an appropriate tool to include both these
features. Thus, in the RHB theory the advantages of a relativistic
description of the RMF approach in the Hartree channel
are combined with that of a finite-range pairing force.

The RMF Lagrangian which describes the nucleons as Dirac spinors 
moving in meson fields is given by \cite{Serot.86}
\begin{eqnarray}
{\cal L}&=& \bar\psi \left( \rlap{/}p - g_\omega\rlap{/}\omega -
g_\rho\rlap{/}\vec\rho\vec\tau - \frac{1}{2}e(1 - \tau_3)\rlap{\,/}A -
g_\sigma\sigma - M_N\right)\psi\nonumber\\
&&+\frac{1}{2}\partial_\mu\sigma\partial^\mu\sigma-U(\sigma)
-\frac{1}{4}\Omega_{\mu\nu}\Omega^{\mu\nu}+ \frac{1}{2}
m^2_\omega\omega_\mu\omega^\mu\\ &&+\frac{1}{2}g_4(\omega_\mu\omega^\mu)^2
-\frac{1}{4}\vec R_{\mu\nu}\vec R^{\mu\nu}+
\frac{1}{2} m^2_\rho\vec\rho_\mu\vec\rho^\mu -\frac{1}{4}F_{\mu\nu}F^{\mu\nu}
\nonumber
\end{eqnarray}
where $M_N$ is the bare nucleon mass and $\psi$ is its Dirac spinor. In 
addition, we have the scalar meson ($\sigma$), isoscalar vector mesons 
($\omega$), isovector vector mesons ($\rho$) and the photons $A^\mu$, 
with the masses $m_\sigma$, $m_\omega$ and $m_\rho$ and the coupling 
constants $g_\sigma$, $g_\omega$, $g_\rho$, respectively. The field 
tensors for the vector mesons are given as 
$\Omega_{\mu\nu}=\partial_\mu\omega_\nu-\partial_\nu\omega_\mu$ 
and by similar expressions for the $\rho$-meson and 
the photon. For a realistic description of nuclear properties a nonlinear
self-coupling $U(\sigma) = \frac{1}{2} m^2_\sigma \sigma^2_{} +
\frac{1}{3}g_2\sigma^3_{} + \frac{1}{4}g_3\sigma^4_{}$ for  
$\sigma$-mesons is taken. We have added the non-linear vector 
self-coupling of $\omega$-meson \cite{Bod.91} as represented by the 
coupling constant $g_4$. 

Using Green's function techniques \cite{Go.58} it has been shown in Ref.
\cite{KR.91} that a relativistic Hartree-Bogoliubov theory can be
implemented using such a Lagrangian. Neglecting retardation effects one 
obtains a relativistic Dirac-Hartree-Bogoliubov (RHB) equations
\begin{equation}
\left(\begin{array}{cc} h & \Delta \\ -\Delta^* & -h^* \end{array}\right)
\left(\begin{array}{r} U \\ V\end{array}\right)_k~=~
E_k\,\left(\begin{array}{r} U \\ V\end{array}\right)_k,
\label{RHB} 
\end{equation}
where $E_k$ are quasiparticle energies and the coefficients $U_k$ and 
$V_k$ are four-dimensional Dirac spinors normalized as
\begin{equation}
\int ( U^+_k U^{}_{k'}~+~V^+_kV^{}_{k'}\, ) d^3r~=~\delta_{kk'}.
\end{equation} 
The average field
\begin{equation}
h~=~\mbox{\boldmath $\alpha p$}~+~g_\omega\omega~+~ \beta(M+g_\sigma
\sigma)~-~\lambda
\label{h-field}
\end{equation}
contains the chemical potential $\lambda$ which is adjusted to the proper 
particle number. The meson fields $\sigma$ and $\omega$ are determined 
self-consistently from the Klein Gordon equations:
\begin{eqnarray}
\left\{-\Delta+m^2_\sigma\right\}\sigma&=& -g_\sigma \rho_s~-~g_2
\sigma^2~-~g_3\sigma^3,\\ 
\left\{-\Delta+m^2_\omega\right\}\omega&=& g_\omega \rho_v~+~g_4\omega^3,
\label{KG}
\end{eqnarray}
with the scalar density $\rho_s=\sum_k \bar V^{}_kV^{}_k$ and the baryon
density $\rho_v=\sum_k V^+_kV^{}_k$.  The sum on $k$ runs only over all the
particle states in the no-sea approximation. The pairing potential 
$\Delta$ in Eq. (\ref{RHB}) is given by
\begin{equation}
\Delta_{ab}~=~\frac{1}{2}\sum_{cd} V^{pp}_{abcd} \kappa_{cd}
\label{gap}
\end{equation}
The RHB equations (\ref{RHB}) are a set of four coupled 
integro-differential equations for the Dirac spinors $U(r)$ and 
$V(r)$ which are obtained self-consistently. Here the RHB calculations 
have been performed by expanding fermionic and bosonic wavefunctions 
in 20 oscillator shells for a spherical configuration. We have used the 
force NL-SH \cite{SNR.93} with the non-linear scalar self-coupling and 
the forces NL-SV1 and NL-SV2 with both the scalar and the vector
self-couplings. The forces NL-SV1 and NL-SV2 have been developed
with a view to soften the high-density equation of state of
nuclear matter and are shown to improve the ground-state properties
of nuclei \cite{SFM.00,Sha.00}. For the pairing channel, we have taken
the finite-range Gogny force D1S. It is known to represent the pairing
properties of a large number of finite nuclei appropriately
\cite{Berger.36}.

In our previous work \cite{SFM.00}, we started with the shell effects at 
the stability line, where we examined the role of $\sigma$- and 
$\omega$-meson couplings on the shell effects in Ni and Sn isotopes. 
It was found that the existing nuclear forces based upon the nonlinear 
scalar self-coupling of $\sigma$-meson exhibit shell effects 
which were stronger than suggested by the experimental data. In order to 
remedy this problem, the nonlinear vector self-coupling of $\omega$-meson 
in the RHB theory was introduced. Consequently, the experimental data on 
shell effects in nuclei about the stability line were reproduced
well \cite{SFM.00}.

Having established the basis, we consider nuclei about $^{80}$Zn ($N=50$) 
in the present work. $^{80}$Zn is a waiting point nucleus 
and here the r-process path comes closest to the $\beta$-stability. 
The empirical values on the binding energy of $^{80}$Zn and  $^{82}$Zn 
are taken from the compilation by Wapstra and Audi \cite{Audi.95}. These data
provide a unique opportunity to probe the nature of the shell effects
away from the stability line.

Results on $S_{2n}$ values for the Zn isotopes are shown in Fig.~1. 
The kink at $N=50$ shows that NL-SH (scalar self-coupling only) 
exhibits shell effects which are much stronger than the 
experimental data. In comparison, the shell gap with NL-SV2 (both the 
scalar and vector self-couplings) \cite{SFM.00} is reduced as compared to 
NL-SH. It is, however, still larger than the experimental one. 
We show in Fig.~1(b) the $S_{2n}$ values obtained with the force NL-SV1 
\cite{SFM.00}. The slope of the kink shows that the shell gap
with NL-SV1 agrees well with the experimental value. Thus, RHB with 
the force NL-SV1 is able to reproduce the empirical shell effects in 
the waiting point region. A comparison with the HFB+SkP results 
\cite{DFT.84} shows that the shell effects with SkP are strongly
quenched. In contrast, the data show that the shell effects in
the waiting point region are much stronger than those predicted by SkP. 
This is consistent with our earlier conclusion \cite{SFM.00} that the shell 
effects with SkP are quenchend strongly already at the stability line. 
This behaviour is evidently due to its high effective mass $m^* \sim  1$ 
\cite{DFT.84} which suppresses the shell gaps significantly. 

In Fig.~1(c) we compare various predictions of the mass formula Extended 
Thomas-Fermi with Strutinsky Integral (ETF-SI) \cite{Abou.95}.
It is widely used in the abundance calculations of the r-process
nucleosynthesis due to its success in reproducing the binding 
energies of nuclei over a wide range of the periodic table. 
It can be seen from the figure that ETF-SI reproduces the shell 
effects in the waiting point region successfully. However, motivated
by the results of HFB+SkP, a phenomenogical quenching of the shell
effects has recently been introduced 
\cite{Pear.96} in ETF-SI, thus producing a new mass table ETF-SI (Q). 
The results [Fig.~1(c)] of the quenched (Q) mass formula show that the 
shell effects are weakened as compared to the experimental data 
and ETF-SI. We have found several other cases where a good agreement of
ETF-SI with the experimental data is deteriorated in going to ETF-SI(Q).
Thus, the introduction of the quenching in ETF-SI (Q) seems to be 
unnecessary notwithstanding the success of ETF-SI. 

The RHB results with NL-SV1 in Fig.~1(b) provide a confidence to extend 
our formalism to explore the inaccessible region of the neutron drip-line 
about $N=82$. In order to visualize how the shell effects evolve 
as one approaches the continuum, we have chosen the isotopic chains 
of Kr, Sr, Zr and Mo ($Z=36-42$). The $S_{2n}$ values as obtained in RHB 
with the various forces are shown in Fig.~2. For a given force, the shell 
gap at $N=82$ shows a steady decrease in moving from Mo (a) to Kr (d).  
Evidently, nuclei become increasingly unbound and a coupling to
the continuum arises in going to larger neutron to proton ratios.

The results [Fig.~2] with the non-linear scalar coupling (NL-SH)
show a shell gap at $N=82$, which is largest amongst all the forces. 
The strong shell effects with NL-SH were also noted for $N=82$ 
nuclei near the drip line in ref. \cite{SLHR.94}. These results
were based upon the BCS pairing and the shell effects
along the stability line were not taken into account appropriately.
This emphazises the importance of the self-consistent pairing and of
the data available on the shell effects in the known region. 

With the force NL-SV2 with the vector self-coupling of
$\omega$-meson, the shell effects [Fig.~2] are milder
as compared to NL-SH for all the chains. This is similar
to that observed at the stability line \cite{SFM.00} and also in Fig.~1(a). 
The results [Fig.~2] with our benchmark force NL-SV1 show that the shell
gap at $N=82$ is reduced as compared to NL-SV2. This is again similar to 
that observed for the Zn isotopes. However, the shell effects with 
NL-SV1 are still stronger as compared to the HFB+SkP results shown in 
the figure. This is especially true for Mo and Zr isotopes at N=82, 
where the r-process path is assumed to pass through. The quenching 
shown by SkP in this region is as expected. As noted earlier
HFB+SkP is known to exhibit a strong quenching at the stability line 
\cite{SFM.00} and near the waiting-point nucleus $^{80}$Zn as shown 
in Fig.~1(b) above. In comparison, the RHB approach with NL-SV1 
reproduces the shell effects about the stability line 
\cite{SFM.00} as well as in the waiting-point region at $N=50$. 

The shell effects with NL-SV1 become successively weaker (which is true 
for all the forces) as one moves to nuclei with higher isospin such 
as $^{120}$Sr [Fig.~2(c)] and $^{118}$Kr [Fig.~2(d)]. The case of Kr 
isotopes [Fig.~2(d)] deserves a special mention. Except with NL-SH, 
all the other forces show a complete washing out of the shell effects. 
This stems from the fact that for $^{118}$Kr, the Fermi energy 
is very close to the continuum and the nucleus is pushed to the 
very limit of binding. The binding energy of an additional neutron  
is close to zero and consequently the shell gap ceases to exist.
Thus, any semblance of the shell effects for $^{118}$Kr ($N=82$) is 
completely lost. Such nuclei at the drip line ($\lambda_n \sim 0$) 
should be of little interest to the r-process as these nuclei show 
no binding to an additional neutron. However, as the r-process path
passes through $S_n \sim 2-4 $ MeV, it is seen [Figs.~2(a) and 2(b)]
that nuclei which ought to contribute to the r-process show a 
persistence of stronger shell gaps in contrast to that seen for Kr 
and Sr. Figures 2(a) and 2(b) show that with NL-SV1 the shell effects 
for Mo (Z=42) and Zr (Z=40) nuclei at $N=82$ are much stronger than 
with SkP. Shell gaps are expected to be even larger for r-process
nuclei with $Z > 42$. Thus, notwithstanding the fact that NL-SV1 is 
commensurate with the experimental data available in the waiting point 
region, it can be concluded that the shell effects about the r-process 
path remain stronger.

We show in Fig.~3(a) the neutron single-particle levels obtained with
NL-SV1 for $N=80$ nuclei. Our focus is the evolution of the shell gap 
at $N=82$ as one approaches the drip line ($\lambda_n \sim 0$). The 
single-particle levels shown in this figures portray the results of 
Fig.~2. The Fermi energy (shown by the dashed lines) approaches the continuum
as one moves towards a larger neutron to proton ratio. The shell gap ($N=82$)
shows a constant decrease as the last neutrons become more and more 
unbound in going from Mo to Kr. For $^{118}$Kr ($N=82$) the Fermi energy 
is close to zero and the shell gap diminishes significantly. 
However, for nuclei on the r-process path such as Mo and Zr, shell
gaps do remain large.

The difference in the response of the various forces near the drip line is 
illustrated in single-particle levels shown for $^{120}$Zr in Fig.~3(b). 
The shell gap at $N=82$ is largest with NL-SH and shows a reduction in 
going to NL-SV1.  On the other hand, the shell gap with SkP is reduced
significantly as compared to NL-SV1. The reduced shell strength with
SkP seems to be generic as discussed above.

In conclusion, we have reproduced the shell effects in the waiting-point 
nucleus $^{80}$Zn with the vector self-coupling of $\omega$-meson in the 
RHB theory. Having our working basis established, we show that the shell 
effects near the r-process path about $N=82$ remain strong vis-a-vis a 
quenching exhibited by HFB+SkP. The oft-discussed quenching with 
SkP is, however, not supported by the available experimental 
data in the waiting-point region. It is noteworthy that on the basis
of the results with SkP, a quenching has been requested for an improved
fit to the global r-process abundances \cite{Kratz.93}. Since SkP fails to
reproduce the shell effects at the stability line and in the waiting-point 
region, an improved fit alone does not necessarily imply a shell quenching 
near the r-process path. On the other hand, our results on the stronger
shell effects at the r-process path are consistent with the
point made in refs. \cite{Hax.97,Qian.97} that neutrino-induced reaction
during a core-collapse supernova can be effective in breaking through
the waiting-point nuclei and that it is not necessary to invoke a
quenching of the shell strength. In view of this, the present results support
neutrino-induced reaction as a plausible mechanism of nucleosynthesis
of heavy nuclei.

This work is supported by the Research Administration Project No. SP056 
of the Kuwait University. We thank Prof. J.M. Pearson for providing us 
ETF-SI(Q) table before its publication. 



\end{multicols}

\addvspace{16mm}

\newpage
\begin{itemize}

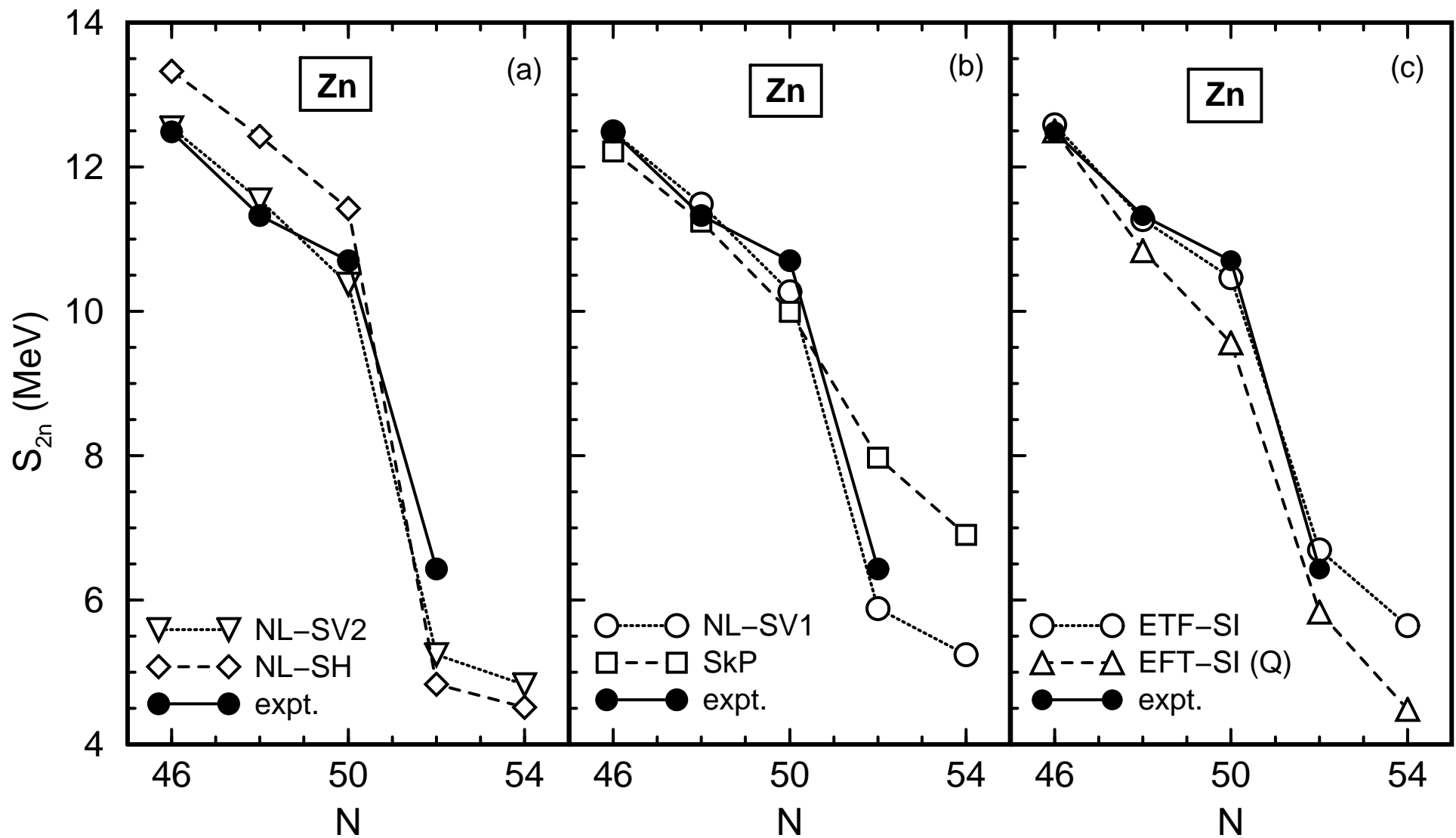
\item[Fig. 1] The $S_{2n}$ values for Zn isotopes in the 
waiting-point region with (a) NL-SH and NL-SV2 (b) NL-SV1 and HFB+SkP 
and (c) the mass formulae ETF-SI, and ETF-SI(Q) with shell quenching, 
as compared to the experimental data.

\item[Fig. 2] Evolution of the shell effects at $N=82$ with 
the various forces for the isotopes chains of Mo, Zr, Sr and Kr in the 
vicinity of the neutron drip line.

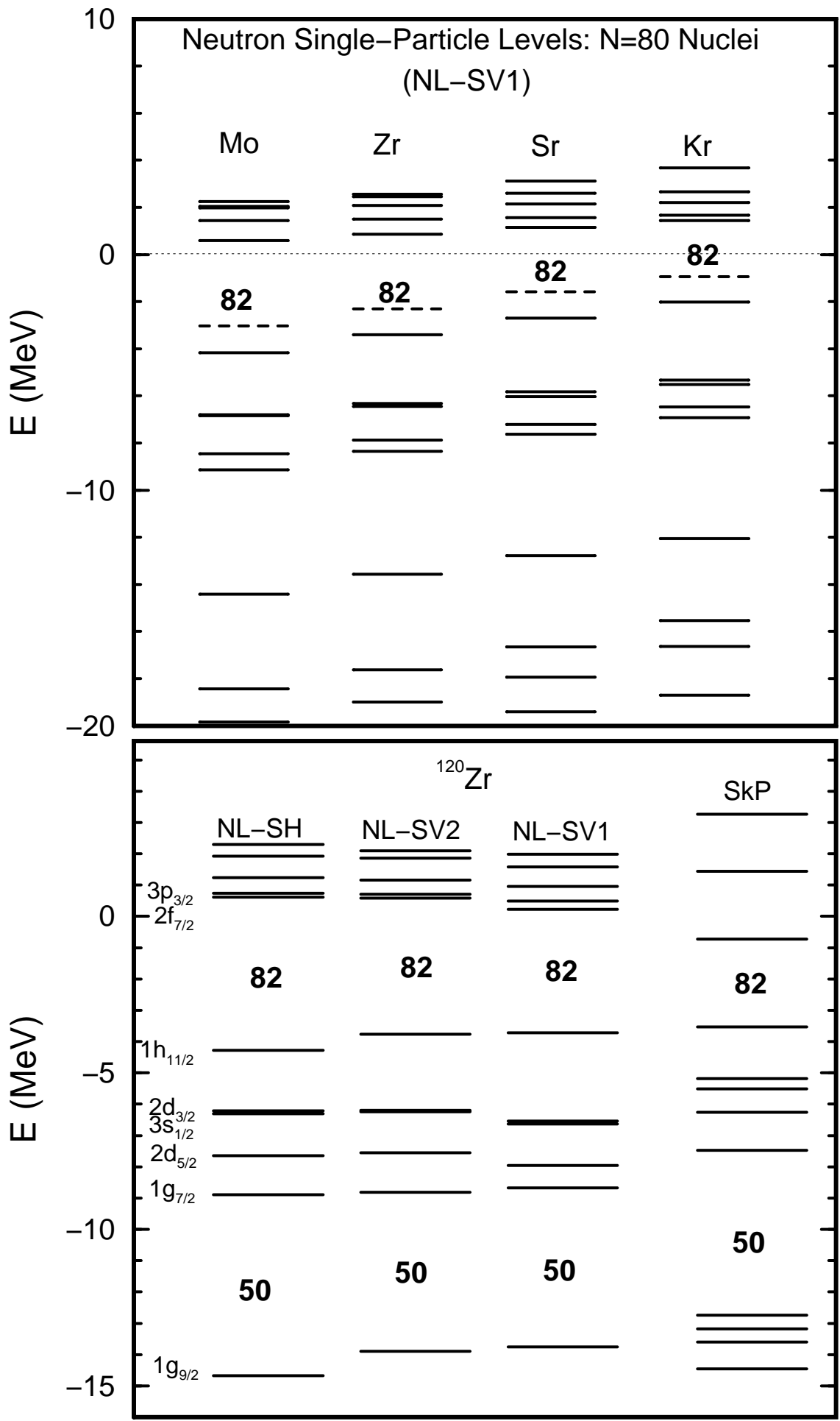
\item[Fig. 3] (a) The neutron single-particle levels for Mo, Zr, Sr and 
Kr nuclei (N=80) near the neutron drip-line with the force NL-SV1. 
The Fermi energy is shown by the dashed lines. (b) The $N=82$ shell 
gap in the neutron single-particle levels for $^{120}$Zr 
with the various forces.

\end{itemize}


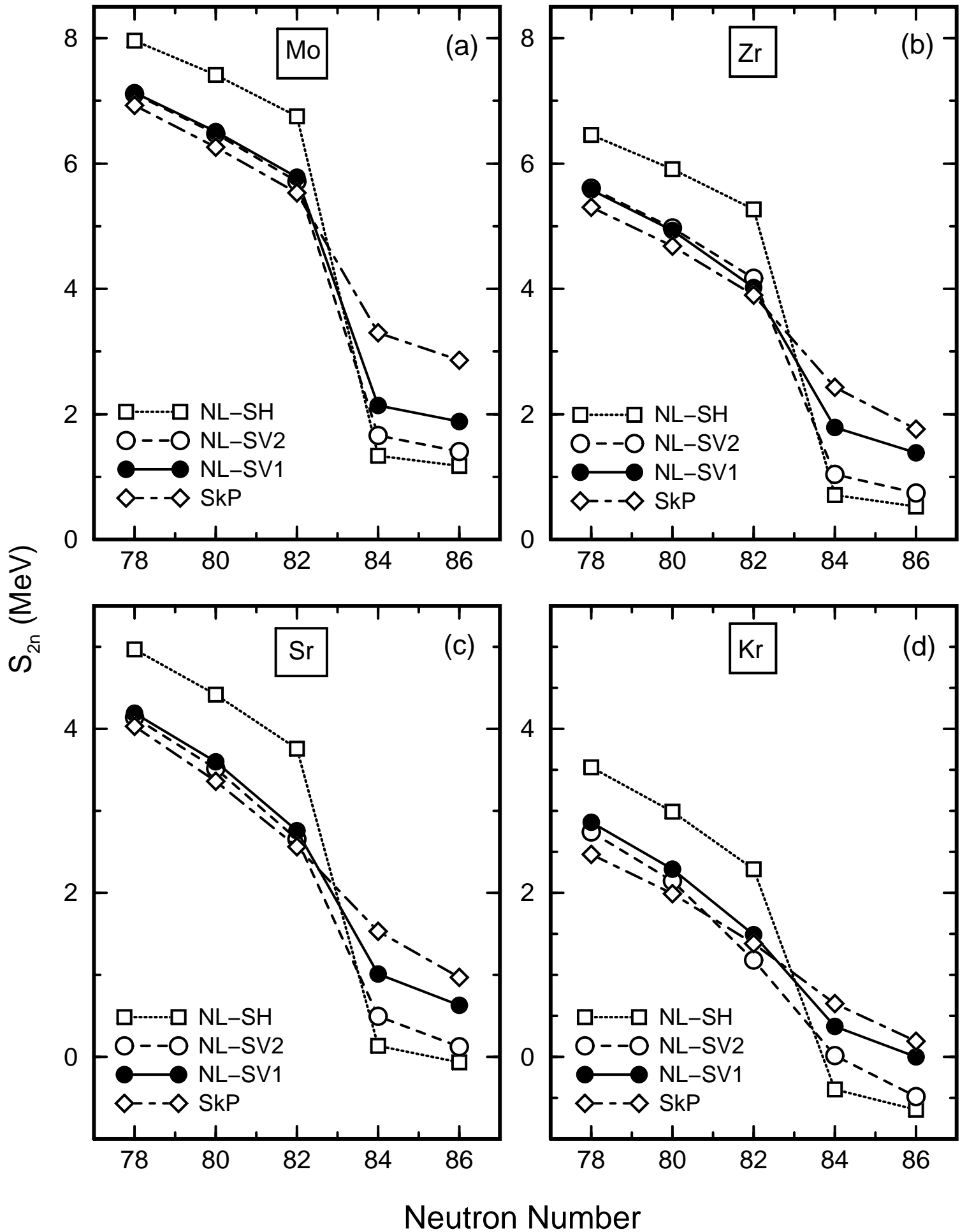
\end{document}